\documentclass[showpacs,aps,prstab,amsmath,amssymb]{revtex4-1}
\usepackage[latin9]{inputenc}
\usepackage{color}
\usepackage{mathrsfs}
\usepackage{amsmath}
\usepackage{graphicx}
\usepackage{esint}

\makeatletter

\DeclareTextSymbolDefault{\textquotedbl}{T1}
\providecommand{\tabularnewline}{\\}

\usepackage{bm}
\usepackage{booktabs}
\usepackage{color}
\usepackage{hyphenat}

\renewcommand{\[}{\begin{equation}}
\renewcommand{\]}{\end{equation}}
\makeatother

\begin{document}
\title{Feasibility study of a hard x-ray FEL oscillator at 3 to 4 GeV based
on harmonic lasing and transverse gradient undulator}
\author{Li Hua Yu, Victor Smaluk, Timur Shaftan, Ganesh Tiwari, Xi Yang}
\address{NSLSII, Brookhaven National Laboratory, NY11973}
\date{3/2/2023}
\begin{abstract}
We studied the feasibility of a hard x-ray FEL oscillator (XFELO)
based on a 3 to 4 GeV storage ring considered for the low-emittance
upgrade of NSLS-II. We present a more detailed derivation of a formula
for the small-gain gain calculation for 3 GeV XFELO published in the
proceedings of IPAC'21 \citep{ipac21yu}. We modified the small-signal
low-gain formula developed by K.J. Kim, et.al. \citep{lindberg,ysLi,kim4}
so that the gain can be derived without taking the \textquotedblleft no
focusing approximation\textquotedblright{} and a strong focusing can
be applied. In this formula, the gain is cast in the form of a product
of two factors with one of them depending only on the harmonic number,
undulator period, and gap. Using this factor, we show that it is favorable
to use harmonic lasing to achieve hard x-ray FEL working in the small-signal
low-gain regime with the medium-energy electron beam (3-4 GeV). Our
formula also allows FEL optimization by varying the vertical gradient
of the undulator, the vertical dispersion, and the horizontal and
vertical focusing, independently. Since a quite high peak current
is required for the FEL, the collective effects of beam dynamics in
medium-energy synchrotrons significantly affect the electron beam
parameters. We carried out a multiple-parameter optimization taking
collective effects into account and the result indicates the XFELO
is feasible for storage ring energy as low as 3 GeV, with local correction
of betatron coupling.
\end{abstract}
\maketitle

\section{Introduction}

An x-ray FEL oscillator based on a transverse gradient undulator (TGU)
\citep{tgu1,tgu2} considered by APS and SLAC collaboration \citep{lindberg,ysLi}
provides a promising direction for a storage-ring-based fully coherent
hard x-ray source. The difficulty associated with the relatively large
energy spread of $10^{-3}$in the storage ring is mitigated by introducing
TGU and dispersion in the FEL by a trade-off with increased transverse
beam size.

The examples in these literatures use electron beam energy of 6 GeV.
For a medium energy light source such as NSLSII, with energy at 3GeV,
the first difficulty is the relatively lower energy. To achieve hard
x-ray at 0.12nm, to satisfy the resonance condition, the required
undulator period for such low energy is less than 1cm, and the gap
of a few mm makes it very difficult to allow the electron beam to
achieve the required beam quality. We are obliged to consider harmonic
lasing under this circumstance.

The harmonic generation and harmonic lasing, for high gain FEL, are
analyzed, for example, in \citep{huang,yurkov}. The harmonic lasing
in a low gain regime is expected to have a slightly different but
similar scaling relation with the harmonic number. However, we need
more quantitative analysis of the scaling relation, particularly because
we are considering the case of lower energy and large energy spread.
We adopt the approach in \citep{lindberg,ysLi} for the low gain formula
which is based on the low gain formula derived by K.J. Kim\citep{kim4}.
For this purpose, we need to follow through with the derivation to
explicitly allow for harmonic lasing. Before going into more detailed
analysis, we first consider the gain formula in 1D Madey theorem with
harmonic number $h$ and when energy spread is negligible, the gain
can be cast in a form convenient for scaling the gain with harmonic
number $h$ and undulator period $\lambda_{u}$:

\begin{equation}
G_{1D}=\left(\frac{hK^{2}[JJ]_{h}^{2}}{\lambda_{u}}\right)\left(\pi^{2}\frac{I}{I_{A}}\frac{L_{u}^{3}}{\gamma^{3}\Sigma}\right)\left(\frac{d}{d\Phi}\left(\frac{\text{sin}\Phi}{\Phi}\right)^{2}\right)\label{eq:G1D}
\end{equation}

where $K$ is the undulator parameter given by the peak field $B_{peak}$
in the resonance condition

\begin{equation}
\lambda_{s}=\frac{\lambda_{u}}{2\gamma_{0}^{2}h}(1+\frac{K^{2}}{2})\label{eq:resonance condition}
\end{equation}

$\gamma_{0}$ is the resonant electron beam energy in the unit of
electron rest mass, $\lambda_{s}$ is the FEL wavelength, $L_{u}=N_{u}\lambda_{u}$
is the undulator length with the number of period $N_{u}$. $[JJ]_{h}=J_{(h-1)/2}\left(\frac{hK^{2}}{4+2K^{2}}\right)-J_{(h+1)/2}\left(\frac{hK^{2}}{4+2K^{2}}\right)$
is the Bessel factor, $\Sigma=2\pi\sigma_{x}\sigma_{y}$ is the electron
beam cross-section area with $\sigma_{x},\sigma_{y}$ the RMS beam
size, $I$ is beam peak current. $I_{A}=4\pi mc^{3}\epsilon_{0}/e\approx17kA$
is the Alfven current. $\Phi=\pi\Delta\nu N_{u}-2\eta h\pi N_{u}$
is the phase advance in the undulator due to detuning, with $\eta=(\gamma-\gamma_{0})/\gamma_{0}$
is the relative energy detuning of mean energy $\gamma$ from resonance,
$\Delta\nu=h(\omega-\omega_{s})/\omega_{s}$ is the laser frequency
detuning from resonance frequency $\omega_{s}=2\pi/\lambda_{s}$ with
harmonic number $h$.

The effect of the energy spread can be obtained by an average of the
3rd factor $\frac{d}{d\Phi}\left(\frac{\text{sin}\Phi}{\Phi}\right)^{2}$over
the energy spread $\sigma_{\eta}$, which gives the effect of the
spread of $\Phi$.

For a given wavelength $\lambda_{s}$, energy $\gamma$, peak current
$I$, undulator length $L_{u}$, and the electron beam cross-section
$\Sigma$, we consider the scaling relation of the gain given by Eq.(\ref{eq:G1D})
when we vary the harmonic number $h$, and the undulator period $\lambda_{u}$.
The second factor is fixed by these parameters. However, in order
to achieve sufficient gain, the need to increase the harmonic number
$h$ becomes clear only after we calculate the first factor $\left(\frac{K^{2}[JJ]_{h}^{2}h}{\lambda_{u}}\right)$
as a function of $h$ and $\lambda_{u}$ while taking into account
the resonance condition Eq.(\ref{eq:resonance condition} ) and the
relation of undulator parameter $K$ to the undulator period $\lambda_{u}$,
peak field $B_{peak}$ and gap $g$. Assume the relation between $B_{peak}$
and $g$ is given by the K.Halbach formula\citep{halbach}, we have:

\begin{align}
 & K=\frac{e\lambda_{u}B_{peak}}{2\pi mc}=93.43\lambda_{u}B_{peak}\nonumber \\
 & B_{peak}=3.1\exp\left(-5.47\frac{g}{\lambda_{u}}+1.8\left(\frac{g}{\lambda_{u}}\right)^{2}\right)\label{eq:Halbach}
\end{align}

For a given $\lambda_{u}$ and $h$, the resonance condition Eq.(\ref{eq:resonance condition})
determines $K$, then Eq.(\ref{eq:Halbach}) determines $B_{peak}$
and the gap $g$. For 3GeV beam, and for $\lambda_{s}=0.12nm$, we
plot $\frac{K^{2}[JJ]_{h}^{2}h}{\lambda_{u}}$ as function of $\lambda_{u}$for
$h=1,5$ respectively in Fig.1a. We plot the gap $g$ and $K$ respectively
in Fig.1b, and Fig.2a as function of $\lambda_{u}$ for $h=1,5$.
From Fig.1b we see that for $h=1$, the range of period to satisfy
the resonance condition is between 4.5mm to 8.2mm. At 4.5mm, the gap
is zero while at the other limit 8.2 mm, the gap is 5.5mm, but the
$K$ approaches zero in Fig. 2a. For $h=5$, the range satisfying
the resonance condition is above 9.2mm, and gap is above 7mm when
$\lambda_{u}>2.2cm$. The very narrow gap for $h=1$ makes it very
difficult for the x-ray FEL to use $h=1$, and from this point of
view of gap alone we would need to consider possibility for higher
harmonic number.

The main issue here is whether there is sufficient gain with $h>1$,
i.e., whether the gain is larger than the total power loss of the
X-ray cavity. We considered a bow-tie configuration with a 200m roundtrip
and the cavity consists of four crystals and two focusing lenses as
identified in Fig. (3) of Ref. \citep{cavity}. To estimate power
loss and outcoupling through the cavity, we considered a monochromatic
radiation beam at 0.12nm with a Gaussian transverse profile in physical
and angular space expected at the steady-state of XFELO. Since angular
filtering from Bragg-Crystal in the reflecting plane dominates power
loss in the cavity, we stabilized the cavities under consideration
by placing two Be lenses with the same focal length at either side
of the undulator. The first lens after the undulator collimates the
diverging radiation beam ensuring a parallel transverse profile of
the radiation beam while propagating through all crystals before the
second lens.

We further confined crystal choices to symmetrical Bragg reflection
cases for simplicity and convenience. We identified a few Bragg-crystals
allowing us to confine overall power loss in the cavity to 5\%. Thorough
details of the optical cavity analysis will be published elsewhere.

First, we consider the factor $\frac{K^{2}[JJ]_{h}^{2}h}{\lambda_{u}}$
in the 1D gain. The Fig.1a shows the advantage of higher $h$. We
compare two points with same gap of $g=$3.8mm, the first factor is
8.4 for $h=1,$while it is 21.2 for $h=5$, even though this gap is
too narrow to be practical. Another pair of points have the same gain
factor $\frac{K^{2}[JJ]_{h}^{2}h}{\lambda_{u}}=8.4$, the gap is 3.8mm
for $h=1,$while for $h=5$ the gap is 6.1mm. The examples show for
the contribution to the first factor, for the same gap, higher harmonic
number has higher gain, while for the same gain, higher harmonic number
has larger gap. The magent point a is the working point we use in
the following sections.

\begin{figure}[t]
\includegraphics[width=0.48\textwidth]{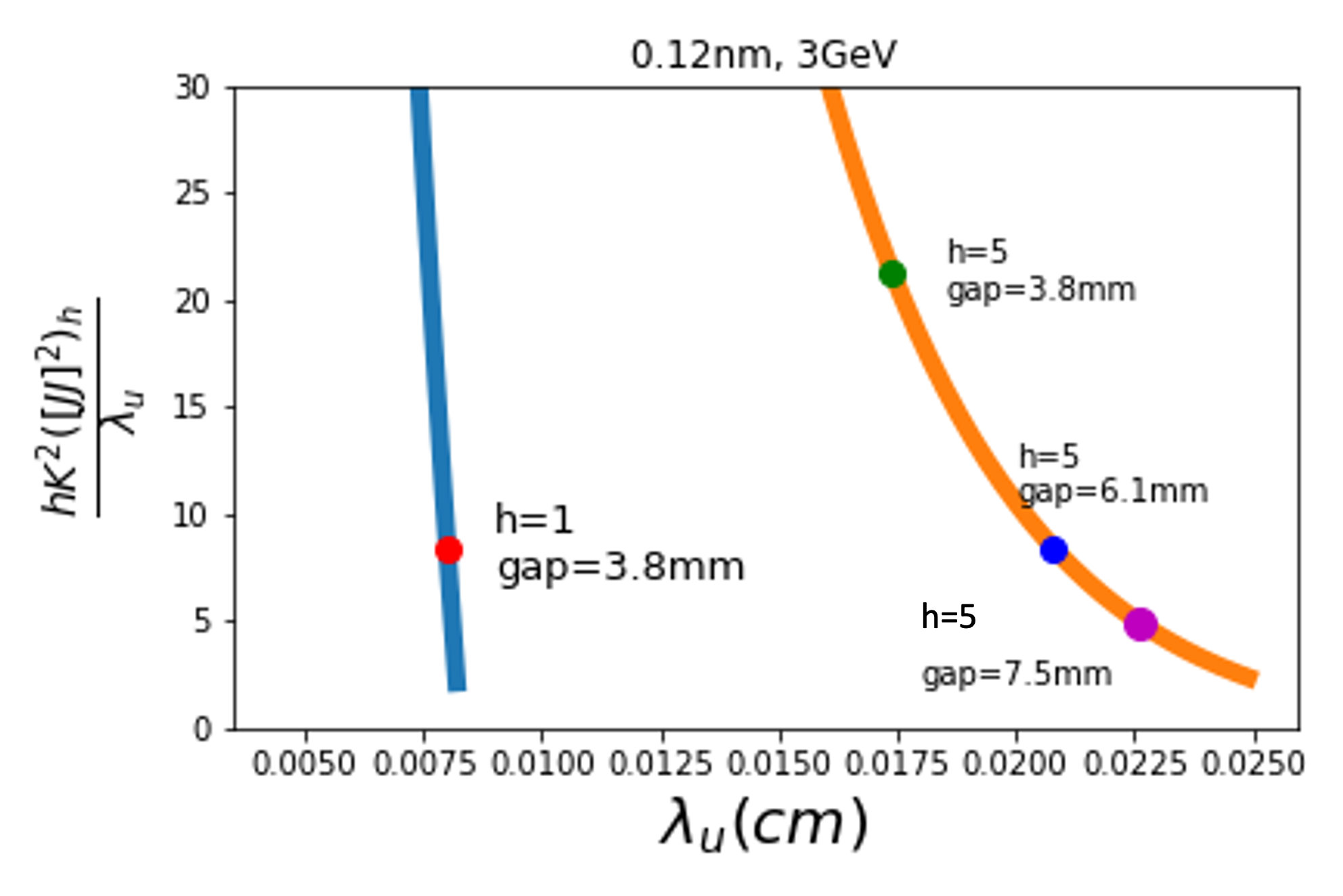}\vspace{-0.0em}\includegraphics[width=0.48\textwidth]{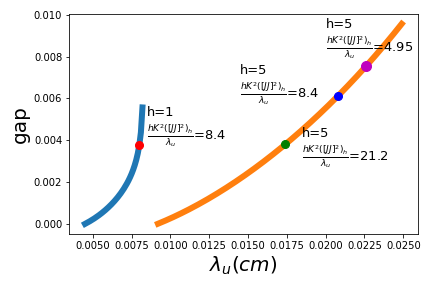}\vspace{-0.0em}

\caption{a:\ $\frac{K^{2}[JJ]_{n}^{2}h}{\lambda_{u}}$ vs. $\lambda_{u}$,
compare h=5 with h=1\ \ \ b:\ gap vs. $\lambda_{u}$,}
\vspace{-0.0em}
\end{figure}

\begin{figure}[t]
\includegraphics[width=0.48\textwidth]{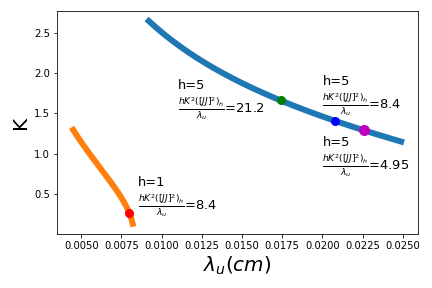}\vspace{-1.0em}\includegraphics[width=0.48\textwidth]{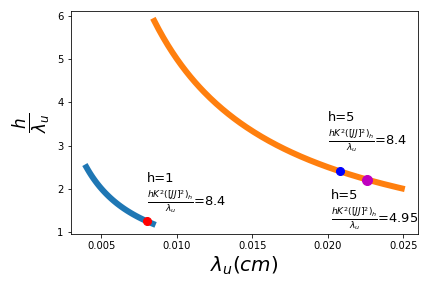}\vspace{-1.0em}\vspace{0.0em}

\caption{a: \ \ K vs. $\lambda_{u}$,\ \ \ b:\ $\frac{h}{\lambda_{u}}$
vs. $\lambda_{u}$,}
\vspace{0.0em}
\end{figure}
However, the contribution from the 3rd factor $\frac{d}{d\Phi}\left(\frac{\text{sin}\Phi}{\Phi}\right)^{2}$
is more complicated, with a maximum of 0.54 at $\Phi=-1.3$, the term
$2\eta h\pi N_{u}=2\eta\pi L_{u}\frac{h}{\lambda_{u}}$ in $\Phi$
has a spread proportional to $h/\lambda_{u}$ due to the energy spread
$\sigma_{\eta}$, the increased spread would reduce the average value
of the 3rd factor. To see its effect on the spread of $\Phi$, we
plot $h/\lambda_{u}$ as a function of $\lambda_{u}$in Fig.2b. For
the pair of points in Fig.1a with the same first gain factor 8.4,
we see the ratio $\frac{h}{\lambda_{u}}$ increases from 1.2 for $h=1$
to 2.6 for $h=5$. This reduction is significant, hence the main question
in the following discussion is whether the gain reduction due to the
large energy spread can be mitigated by TGU and dispersion sufficiently
to maintain the required gain.

In the formulation developed in \citep{lindberg,ysLi,kim4} about
the 3D gain, if we assume the gradient is in the vertical direction,
then the undulator parameter $K=K_{0}(1+\alpha y)$, the energy is
$\gamma=\gamma_{0}(1+\eta)$, for $\alpha y\ll1,\eta\ll1$, the resonance
condition becomes

\begin{align}
 & \lambda=\frac{\lambda_{u}}{2\gamma^{2}h}(1+\frac{K^{2}}{2})\label{eq:resonance condition with gradient}\\
 & =\frac{\lambda_{u}}{2\gamma_{0}^{2}h}\frac{1+\frac{K_{0}^{2}}{2}(1+\alpha y)^{2}}{(1+\eta)^{2}}\approx\frac{\lambda_{u}}{2\gamma_{0}^{2}h}\left(1+\frac{K_{0}^{2}}{2}\right)\left(1+\frac{2K_{0}^{2}}{2+K_{0}^{2}}\alpha y-2\eta\right)\nonumber 
\end{align}

If we assume the dispersion is $D$ in vertical direction, the vertical
distritbution is in gaussian form $\exp(-\frac{(y-D\eta)^{2}}{2\sigma_{y}^{2}})$.
the centroid of the electron beam with energy $\eta$ is shifted to
$D\eta$, then for very small $\sigma_{y}$, we can take $y\approx D\eta$,
and the deviation from resonance condition due to the spread in $\eta$
is given by $\frac{2K_{0}^{2}}{2+K_{0}^{2}}\alpha D\eta-2\eta$. If
$\alpha D=\frac{2+K_{0}^{2}}{K_{0}^{2}}$, then the gain reduction
due to large energy spread is mitigated. However, there is a trade
off with increased verital beam size. The term $\frac{2K_{0}^{2}}{2+K_{0}^{2}}\alpha\sigma_{y}$
causes a deviation from resonance condition in Eq.(\ref{eq:resonance condition with gradient})
and $D\sigma_{y}$ increases the transverse beam cross section $\Sigma$
in the gain formula Eq.(\ref{eq:G1D}). Both effects reduce the gain,
and both are determined by the emittance and focusing in the undulator.

To study the beam quality required for an x-ray FEL working in the
medium energy range, we need to apply the gain formula developed in
\citep{lindberg,ysLi,kim4} for optimization, with the 3D effect of
diffraction, beam divergence, and betatron motion taken into account,
and in particular, with addition to include harmonic lasing. In following
the procedure to optimize the parameters, we realized that in the
formulation in \citep{lindberg,ysLi,kim4}, a ``no focusing'' approximation
is taken in the process of derivation of the gain formula using the
``brightness function'', as explained clearly after the Eq.(24)
of \citep{kim4} by K.J. Kim. The ``no focusing'' approximation
essentially neglects the focusing. In our optimization process, we
often found that we need to increase the focusing in the undulator,
and we reached a set of parameters that violated the condition required
by the ``no focusing'' approximation. Hence it is desirable to develop
an approach to calculate the gain without taking the ``no focusing''
approximation.

In the following we present a derivation of gain formula without taking
the ``no focusing'' approximation, and some examples of the required
parameters for a medium energy storage ring such as NSLSII. In Section
II, we first follow \citep{kim4} to describe the general 3D gain
formula resemble 1-D Madey theorem, the Eq.(23) in \citep{kim4}.
This general gain formula, without being given a specific gaussian
form of the electron beam distribution and the input radiation field,
is our starting point, which is also the last step in \citep{lindberg,ysLi,kim4}
right before taking the ``no focusing'' approximation. Then in Section
III, we consider a focusing lattice interlaced with the undulator,
the system has nearly constant beta and dispersion functions and hence
has nearly constant transverse beam profile in the undulator, this
allows us to carry out a multivariable gaussian integration without
taking the ``no focusing'' approximation and reduce the gain formula
to a double integral, similar to the result of \citep{lindberg,ysLi,kim4}.
In Section IV we present some examples of gain optimization using
the formula to study the required paramters for an X-ray FEL oscillator
in a medium energy storage ring such as the upgrade being considered
at the NSLSLII. In Section V, we carry out gain optimization with
collective effects taken into account.The resut indicates the feasibility
of XFELO for a 3GeV storage ring at NSLSII. In the Appendix I, in
order to clarify the notations in this paper and in particular to
include the harmonic number $h$ in the formulation, we briefly describe
the steps that lead to the general 3D gain formula in Section I, starting
from the combined Maxwell-Vlasov equations.

\section*{II Gain Formula}

The gain in small signal, low gain regime taking into account of the
3D effect of diffracrion, beam divergence and betatron motion, is
given in the following Eq.(\ref{eq:general gain} ), which is the
Eq.(23) of \citep{kim4}, with a minor elaboration of introducing
the transverse gradient and dispersion for TGU as in \citep{lindberg,ysLi}.
For convenience we adopt nearly identical notation as \citep{kim4},
with only few exceptions to unite with the notations in \citep{lindberg,ysLi,kim4}
and the notations we used in the early development of the coupled
Maxwell-Vlasov equations for high gain FEL\citep{wang,prA,PRL,HGHG}.
The gain expression has multiple integrations to be carried out for
application:

\begin{align}
 & G=-c_{h}\frac{\int d\eta d\mathbf{\boldsymbol{x}}d\mathbf{\boldsymbol{p}}\overline{F}(\mathbf{\boldsymbol{x}},\mathbf{\boldsymbol{p}},\eta;0)\frac{\partial}{\partial\eta}\mid\int d\boldsymbol{\phi}A_{\nu}^{(0)}(\boldsymbol{\phi},0)U_{\nu}^{*}(\eta,\boldsymbol{\phi},\mathbf{\boldsymbol{p}},\mathbf{\boldsymbol{x}})\mid^{2}}{\int\mid A^{(0)}(\boldsymbol{\phi})\mid^{2}d\boldsymbol{\phi}}\label{eq:general gain}
\end{align}

where $\mathbf{x}=(x,y)$, $\mathbf{p}=(p_{x},p_{y})$ , $\boldsymbol{\phi}=(\phi_{x},\phi_{y})$.
$\overline{F}$ is the averaged smooth electron beam background distribution,
we assume

\begin{equation}
\overline{F}(\mathbf{\boldsymbol{x}},\mathbf{\boldsymbol{p}},\eta;z)=\frac{e^{-\eta^{2}/(2\sigma_{\eta}^{2})}}{(2\pi)^{3/2}\sigma_{px}\sigma_{py}\sigma_{\eta}}\exp(-\frac{x^{2}}{2\sigma_{x}^{2}}-\frac{(y-D\eta)^{2}}{2\sigma_{y}^{2}})\exp(-\frac{p_{x}^{2}}{2\sigma_{px}^{2}}-\frac{p_{y}^{2}}{2\sigma_{py}^{2}})\label{eq:F}
\end{equation}

So the electron beam density is $n_{0}\int d\eta dp_{x}dp_{y}\overline{F}(x,p,\eta;z)$,
independent of $z$. $n_{0}$ is normalized such that when $D=0$,
i.e., before entering into the dispersion region, the peak density
at $x=y=0$ is $n_{0}$, so $\int d\eta dp_{x}dp_{y}\overline{F}(0,0,p_{x},p_{y},,\eta;0)\mid_{D=0}=1$.
We assume approximately the electron beam in the undulator is matched
with constant betatron function such that $\sigma_{px}=k_{\beta x}\sigma_{x}=\sigma_{x}/\beta_{x}$,
$\sigma_{py}=k_{\beta y}\sigma_{y}=\sigma_{y}/\beta_{y}$, $p_{x}=\frac{dx}{dz}$,$p_{y}=\frac{dy}{dz}$.

$A_{\nu}(\phi_{x},\phi_{y},z)$ is the angular representation of Fourier
transform $a_{\nu}$ of the horizontal electric field component $E$
at frequency $\omega=\nu\omega_{1}$, while the resonance harmonic
frequency is $\omega_{s}=h\omega_{1}$in Eq.(\ref{eq:resonance condition}),
where $\omega_{1}=2\pi/\lambda_{1}$ is the fundamental frequency.

\begin{align}
 & E(x,y,z,t)=\frac{1}{\sqrt{2\pi}}\int e^{-i\Delta\nu k_{u}z}e^{-i\nu\left(k_{1}z-\omega_{1}t\right)}a_{\nu}(x,y,z)\omega_{1}d\nu\label{eq:Aanular}\\
 & A_{\nu}(\phi_{x},\phi_{y};z)=\frac{1}{\lambda^{2}}\intop d\mathbf{x}a_{\nu}(\mathbf{x},z)e^{ik\left(x\phi_{x}+y\phi_{y}\right)}\nonumber 
\end{align}

where $\omega=2\pi/\lambda,k_{u}=2\pi/\lambda_{u},k_{1}=\omega_{1}/c,k=\omega/c=\nu\omega_{1}/c=\nu k_{1}$,
the detuning is given by $\left(\omega-\omega_{s}\right)/\omega_{s}=(\nu-h)/h=\Delta\nu/h$.

We assume the input radiation $a_{\nu}^{(0)}(x,y,z)$, the solution
of the Maxwell equation in free space, is a gaussian beam with frequency
$\nu$ and Rayleigh range $z_{Rx}=\frac{\pi w_{x}^{2}}{\lambda},z_{Ry}=\frac{\pi w_{x}^{2}}{\lambda}$,
where $w_{x}=2\sigma_{rx}$,$w_{y}=2\sigma_{ry}$. $\sigma_{rx},\sigma_{ry}$
are the RMS input laser beam size. Since in the gain formula Eq.(\ref{eq:general gain})
the factor $A_{\nu}^{(0)}A_{\nu}^{*(0)}$ appears in both the numerator
and the denominator, to simplify writing, we shall take the constant
coefficients as $1$ in $a_{\nu}^{(0)}(x,y,z)$ without changing the
gain $G$.

\begin{align*}
 & E(x,z)\sim a_{\nu}^{(0)}(x,y,z)\exp(-ikz)=\frac{1}{\sqrt{q_{x}(z)}}\frac{1}{\sqrt{q_{y}(z)}}\exp(-ik\frac{x^{2}}{2q_{x}(z)}-ik\frac{y^{2}}{2q_{y}(z)}-ikz)\\
 & q_{x}(z)=z+iz_{Rx},q_{y}(z)=z+iz_{Ry}
\end{align*}

Then Eq.(\ref{eq:Aanular}) gives the angular representation of input
radiation $A_{\nu}^{(0)}(\phi_{x},\phi_{y},0)$ in Eq.(\ref{eq:general gain}):

\begin{align}
 & A_{\nu}^{(0)}(\phi_{x},\phi_{y},z)=\frac{1}{\sqrt{\lambda}}\exp(ik(z+iz_{Rx})\frac{\phi_{x}{}^{2}}{2})\frac{1}{\sqrt{\lambda}}\exp(ik(z+iz_{Ry})\frac{\phi_{y}{}^{2}}{2})=A_{\nu}^{(0)}(\phi_{x},\phi_{y},0)\exp(ikz\frac{\phi_{x}{}^{2}+\phi_{y}{}^{2}}{2})\label{eq:Anu0}\\
 & \text{with }A_{\nu}^{(0)}(\phi_{x},\phi_{y},0)=\frac{1}{\lambda}\exp(-kz_{Rx}\frac{\phi_{x}{}^{2}}{2}-kz_{Ry}\frac{\phi_{y}{}^{2}}{2})\nonumber 
\end{align}

Thus the angular divergence RMS value is $\sigma_{\phi x}=\sigma_{rx}/z_{Rx},\sigma_{\phi y}=\sigma_{ry}/z_{Ry}$
.

The undulator radiation amplitude in Eq.(\ref{eq:general gain}) is
$U_{\nu}^{*}(\eta,\phi_{x},\phi_{y},p_{x},y)$

\begin{align}
 & U_{\nu}^{*}(\boldsymbol{\phi},\eta,\mathbf{x},\mathbf{p};z)=\intop_{-L/2}^{L/2}dse^{-ik\mathbf{x}\boldsymbol{\phi}}e^{i\intop_{0}^{s}ds_{1}\xi_{\nu}(\boldsymbol{\phi},\eta,\mathbf{x},\mathbf{p};s_{1})}\nonumber \\
 & \text{where }\xi_{\nu}(\phi,\eta,\mathbf{x},\mathbf{p};s)=((\Delta\nu-2\nu\eta)k_{u}+\nu\frac{2K_{0}^{2}}{2+K_{0}^{2}}\alpha k_{u}y_{0}+\frac{k}{2}((p_{0x}-\phi_{x})^{2}+k_{\beta x}^{2}x_{0}^{2}+(p_{0y}-\phi_{y})^{2}+k_{\beta y}^{2}y_{0}^{2})\label{eq:Udef}
\end{align}

where $\mathbf{x}_{0}(s;\mathbf{x},\mathbf{p})$,$\mathbf{p}_{0}(s;\mathbf{x},\mathbf{p})$
are solutions of the equations of betatron motion

\begin{align}
\text{} & \frac{dx}{ds}=p_{x},\frac{dy}{ds}=p_{y}\nonumber \\
 & \frac{dp_{x}}{ds}=-k_{\beta x}^{2}x^{2},\frac{dp_{y}}{ds}=-k_{\beta y}^{2}y^{2}\label{eq:betaeq}
\end{align}

with the initial condition $\mathbf{x}_{0}(s=0;\mathbf{x},\mathbf{p})=\mathbf{x}$,$\mathbf{p}_{0}(s=0;\mathbf{x},\mathbf{p})=\mathbf{p}$.
Here we neglect the focusing introduced by the gradient $\alpha$,
and we assumed the focusing comes from the natural focusing and the
external focusing of a FODO lattice, and approximate the beta function
by constants.

The constant $c_{h}=\frac{e^{2}K^{2}[JJ]_{h}^{2}n_{0}}{8mc^{2}\epsilon_{0}\gamma^{3}\lambda^{2}}$,
where the electron beam density at the peak is determined by current
$I=ecn_{0}\Sigma=2\pi\sigma_{x}\sigma_{y}ecn_{0}$ (the current of
an approximately flat top bunch with transverse gaussian distribution).
To write the gain $G$ into a form convenient for scaling with $h$
and $\lambda_{u}$, as shown in Eq.(\ref{eq:G1D}), we have (with
Alven current $I_{A}=4\pi mc^{3}\epsilon_{0}/e)$

\begin{equation}
c_{h}=\frac{I}{I_{A}}\frac{\pi K^{2}[JJ]_{h}^{2}}{\gamma^{3}}\frac{1}{4\pi\sigma_{x}\sigma_{y}\lambda^{2}}\label{eq.ch}
\end{equation}

To calculate the gain $G$ we need to carry out the multivariable
integral in Eq.(\ref{eq:general gain}). Before the multivariable
integration, the approach in\citep{lindberg,ysLi} is to first take
a ``no focusing approximation'' by neglecting the terms $k_{\beta x}^{2}x_{0}^{2}+k_{\beta y}^{2}y_{0}^{2}$,
following the step prescribed in \citep{kim4}. With this approximation,
a gain formula can be transformed into a form where the integrand
becomes a convolution of the distribution function with the radiation
brightness and undulator brightness. This form is appropriate for
a gaussian integration which finally leads to a double integral convenient
for numerical calculation.

The condition for neglecting $k_{\beta x}^{2}x_{0}^{2}$ in $(p_{0x}-\phi_{x})^{2}+k_{\beta x}^{2}x_{0}^{2}$,
because $p_{x}$ and $k_{\beta x}x$ are about the order of $k_{\beta x}\sigma_{x}$,
corresponds to require $\sigma_{\phi x}\gg k_{\beta x}\sigma_{x}$.
Same way it requires $\sigma_{\phi y}\gg k_{\beta y}\sigma_{y}$.
We found often our optimization leads to a set of parameters that
violate this condition, in particular, when emittance is not very
small and we need to increase $k_{\beta}$. In fact, in the example
we developed in Section IV, for $\epsilon_{x0}=80pm,$the optimized
$\beta_{x}=6.6m$ and Rayleigh range $Z_{Rx}=4m$ gives $k_{\beta}\sigma_{x}/\sigma_{\phi x}\approx2.3$.

Hence we would like to try to take into account the effect of the
betatron motion in the gain optimization without the limitation imposed
by this condition. To maintain approximately constant beta and dispersion
functions along the full length of the undulator, we considered a
FODO lattice of focusing interlaced with the sections of the undulator,
with dipole correctors next to the quadrupoles of the FODO lattice
to keep an almost straight line for an approximately constant dispersion
in the transverse gradient undulator. When the beta functions and
dispersion are kept approximately constant, the beam transverse profile
is approximately invariant along the undulator, the interaction between
the effects of the betatron motion, the dispersion, and the FEL embodied
in the integral in Eq.(\ref{eq:general gain}) is significantly simplified.
Under this circumstance, it turns out without neglecting the focusing,
the multiple integral are still possible to be reduced to a double
integral, mainly because the integrations over $x,p_{x},y,p_{y},\eta,\phi_{x},\phi_{y}$
are all gaussian, as given in Section III.

\section*{III Gain Calculation by gaussian integration}

To write the integrand in Eq.(\ref{eq:general gain}) into a gaussian
integral, we first separate the variable $\boldsymbol{\phi}$ in Eq.(\ref{eq:Udef})

\begin{align}
 & \xi_{\nu}(\phi,\eta,\mathbf{x},\mathbf{p};s)=\xi_{\nu}^{(0)}(\eta,x,p;s)-k\mathbf{p}_{0}(s)\boldsymbol{\phi}+\frac{k}{2}\boldsymbol{\phi}^{2}\nonumber \\
 & \text{where }\xi_{\nu}^{(0)}(\eta,x,p;s)\equiv\xi_{\nu}(\phi=0,\eta,\mathbf{x},\mathbf{p};s_{1})=(\Delta\nu-2\nu\eta)k_{u}+\nu\frac{2K_{0}^{2}}{2+K_{0}^{2}}\alpha k_{u}y_{0}+\frac{k}{2}(p_{x}^{2}+k_{\beta x}^{2}x^{2}+p_{y}^{2}+k_{\beta y}^{2}y^{2})\label{eq:ksi0}
\end{align}

Because the last term in $\xi_{\nu}^{(0)}$ is invariant independent
of $s$, $\mathbf{x}_{0}$,$\mathbf{p}_{0}$ are replaced by the initial
value $\mathbf{x},\mathbf{p}$, and the only term dependent on $s$
is $4\nu\zeta_{0}\alpha k_{u}y_{0}(s)$. Since $\mathbf{x}+\intop_{0}^{s}ds_{1}\mathbf{p}_{0}(s)\boldsymbol{\phi}=\mathbf{x_{0}}(s)$,
we have $U_{\nu}^{*}(\boldsymbol{\phi},\eta,\mathbf{x},\mathbf{p};z)=\intop_{-L/2}^{L/2}dse^{-ik\mathbf{x_{0}}(s)\boldsymbol{\phi}+i\frac{k}{2}\boldsymbol{\phi}{}^{2}s}e^{i\intop_{0}^{s}ds_{1}\xi_{\nu}^{(0)}(\eta,\mathbf{x},\mathbf{p};s_{1})}$
in Eq.(\ref{eq:Udef}). So the gain in Eq.(\ref{eq:general gain})
is

\begin{align}
 & G=-c_{h}I_{1e}\left(\int\mid A^{(0)}(\boldsymbol{\phi})\mid^{2}d\boldsymbol{\phi}\right)^{-1}\nonumber \\
 & \text{where }I_{1e}\equiv\int\int\int d\eta d\mathbf{x}d\mathbf{p}\overline{F}(\mathbf{x},\mathbf{p},\eta;0)\frac{\partial}{\partial\eta}\mid\intop_{-L/2}^{L/2}dsI_{1r}e^{i\intop_{0}^{s}ds_{1}\xi_{\nu}^{(0)}(\eta,\mathbf{x},\mathbf{p};s_{1})}\mid^{2}\label{G2}\\
 & \text{with }I_{1r}(s)\equiv\int d\boldsymbol{\phi}A^{(0)}(\boldsymbol{\phi})e^{-ik\mathbf{x_{0}}(s)\boldsymbol{\phi}+i\frac{k}{2}\boldsymbol{\phi}{}^{2}s}\nonumber 
\end{align}

substitute Eq.(\ref{eq:Anu0}) into $I_{1r}(s)$ and use $x_{0}(s)=x\cos(\frac{s}{\beta_{x}})+\beta_{x}p_{x}\sin(\frac{s}{\beta_{x}}),y_{0}(s)=y\cos(\frac{s}{\beta_{y}})+\beta_{y}p_{y}\sin(\frac{s}{\beta_{y}})$,
we find

\begin{align}
 & I_{1r}(s)=-i\frac{1}{\sqrt{z_{Rx}-is}\sqrt{z_{Ry}-is}}\exp(c_{x}(s)x^{2}+2c_{xp}(s)xp_{x}+c_{px}(s)p_{x}^{2})\exp(c_{y}(s)y^{2}+2c_{yp}(s)yp_{y}+c_{py}(s)p_{y}^{2})\nonumber \\
 & \text{where }c_{x}(s)\equiv-k\frac{\cos^{2}(\frac{s}{\beta_{x}})}{2(z_{Rx}-is)},c_{px}(s)\equiv-k\frac{\beta_{x}^{2}\sin^{2}(\frac{s}{\beta_{x}})}{2(z_{Rx}-is)},c_{xp}(s)\equiv-k\frac{\beta_{x}\cos(\frac{s}{\beta_{x}})\sin(\frac{s}{\beta_{x}})}{2(z_{Rx}-is)}\label{eq:Ireq}\\
 & c_{y}(s)\equiv-k\frac{\cos^{2}(\frac{s}{\beta_{y}})}{2(z_{Ry}-is)},c_{py}(s)\equiv-k\frac{\beta_{y}^{2}\sin^{2}(\frac{s}{\beta_{y}})}{2(z_{Ry}-is)},c_{yp}(s)\equiv-k\frac{\beta_{y}\cos(\frac{s}{\beta_{y}})\sin(\frac{s}{\beta_{y}})}{2(z_{Ry}-is)}\nonumber \\
 & \left(\int\mid A^{(0)}(\boldsymbol{\phi})\mid^{2}d\boldsymbol{\phi}\right)^{-1}=2\pi\text{\ensuremath{w_{x}w_{y}}}\nonumber \\
 & I_{1e}=\int\int\int d\eta d\mathbf{x}d\mathbf{p}\overline{F}(\mathbf{x},\mathbf{p},\eta;0)\frac{\partial}{\partial\eta}\intop_{-L/2}^{L/2}\intop_{-L/2}^{L/2}dsdz\mid I_{1r}(s)I_{1r}^{*}(z)\mid e^{i\intop_{0}^{s}ds_{1}\xi_{\nu}^{(0)}(\eta,x,p;s_{1})}e^{-i\intop_{0}^{z}ds_{1}\xi_{\nu}^{(0)}(\eta,x,p;s_{1})}\nonumber 
\end{align}

Substitute $\intop_{0}^{s}ds_{1}y_{0}(s)$ into Eq.(\ref{eq:ksi0}),
we get

\begin{align}
 & \intop_{0}^{s}ds_{1}\xi_{\nu}^{(0)}(\eta,x,p;s_{1})=(\Delta\nu-2\nu\eta)k_{u}s+\frac{k}{2}(p_{x}^{2}+k_{\beta x}^{2}x^{2}+p_{y}^{2}+k_{\beta y}^{2}y^{2})s\nonumber \\
 & +\nu\frac{2K_{0}^{2}}{2+K_{0}^{2}}\alpha k_{u}\beta_{y}y\left(\sin(\frac{s}{\beta_{y}})-\sin(\frac{z}{\beta_{y}})\right)-\nu\frac{2K_{0}^{2}}{2+K_{0}^{2}}\alpha k_{u}\beta_{y}^{2}p_{y}\left(\cos(\frac{s}{\beta_{y}})-\cos(\frac{z}{\beta_{y}})\right)\label{eq:intksi}
\end{align}

With$\text{ }\frac{\partial}{\partial\eta}i\intop_{0}^{s}ds_{1}\xi_{\nu}^{(0)}(\eta,x,p,s_{1})-\frac{\partial}{\partial\eta}i\intop_{0}^{z}ds_{1}\xi_{\nu}^{(0)}(\eta,x,p,s_{1})=-2i\nu k_{u}(s-z)$,
we have

\begin{align*}
 & I_{1e}=\intop_{-L/2}^{L/2}\intop_{-L/2}^{L/2}dsdz\int\int\int d\eta d\mathbf{x}d\mathbf{p}I_{1r}(s)I_{1r}^{*}(z)\overline{F}(\mathbf{x},\mathbf{p},\eta;0)e^{i\intop_{0}^{s}ds_{1}\xi_{\nu}^{(0)}(\eta,x,p;s_{1})}e^{-i\intop_{0}^{z}ds_{1}\xi_{\nu}^{(0)}(\eta,x,p;s_{1})}(-2i\nu k_{u})(s-z)
\end{align*}

Collect all the exponential factors in $I_{1r}(s)I_{1r}^{*}(z)$ $\overline{F}(x,p,\eta;0)$
and in $e^{i\intop_{0}^{s}ds_{1}\xi_{\nu}^{(0)}(\eta,x,p;s_{1})},e^{-i\intop_{0}^{z}ds_{1}\xi_{\nu}^{(0)}(\eta,x,p;s_{1})}$
together (See Eq.(\ref{eq:Ireq}), and Eq.(\ref{eq:F}), and Eq.(\ref{eq:intksi}))
and let

\begin{equation}
D_{R}(s,z)\equiv\sqrt{z_{Rx}-is}\sqrt{z_{Ry}-is}\sqrt{z_{Rx}+iz}\sqrt{z_{Ry}+iz}\label{eq:DR}
\end{equation}

we find

\begin{align}
 & I_{1e}=\intop_{-L/2}^{L/2}\intop_{-L/2}^{L/2}dsdz\left(\frac{(-i2\nu k_{u})(s-z)}{D_{R}(s,z)}\frac{\exp(i\Delta\nu k_{u}(s-z))}{(2\pi)^{3/2}\sigma_{px}\sigma_{py}\sigma_{\eta}}\right)I_{x}(s,z)I_{y\eta}(s,z)\label{eq:I1e}\\
 & \text{where }I_{x}(s,z)=\int\int dxdp_{x}\exp(-\Phi_{x})\nonumber \\
 & I_{y\eta}(s,z)=\int\int d\eta dydp_{y}\exp(-\Phi_{y\eta})\nonumber 
\end{align}

with

\begin{align}
 & \Phi_{x}=A_{x}x^{2}+A_{px}p_{x}^{2}+B_{xp}xp_{x}\nonumber \\
 & A_{x}=\frac{1}{2\sigma_{x}^{2}}D_{x},D_{x}=1-2\sigma_{x}^{2}\left[c_{x}(s)+c_{x}^{*}(z)\right]-ikk_{\beta x}^{2}\sigma_{x}^{2}(s-z)\label{eq:phix}\\
 & A_{px}=\frac{1}{2\sigma_{px}^{2}}D_{px},D_{px}=1-2\sigma_{px}^{2}\left[c_{px}(s)+c_{px}^{*}(z)\right]-ik\sigma_{px}^{2}(s-z)\nonumber \\
 & B_{xp}=-2\left[c_{xp}(s)+c_{xp}^{*}(z)\right]\nonumber 
\end{align}

and

\begin{align}
 & \Phi_{y\eta}=A_{\eta}\eta^{2}+A_{y}y^{2}+A_{py}p_{y}^{2}+B_{\eta y}y\eta+B_{yp}y\eta+B_{\eta}\eta+B_{y}y+B_{py}p_{y}\nonumber \\
 & A_{\eta}\equiv\frac{1}{2\sigma_{\eta}^{2}}+\frac{D^{2}}{2\sigma_{y}^{2}},A_{py}\equiv\frac{1}{2\sigma_{py}^{2}}D_{py},D_{py}=1-2\sigma_{py}^{2}\left[c_{py}(s)+c_{py}^{*}(z)\right]-ik\sigma_{py}^{2}(s-z)\label{eq:phietay}\\
 & A_{y}\equiv\frac{1}{2\sigma_{y}^{2}}-\left[c_{y}(s)+c_{y}^{*}(z)\right]-i\frac{k}{2}k_{\beta y}^{2}(s-z)\nonumber \\
 & B_{\eta y}=-\frac{D}{\sigma_{y}^{2}},B_{yp}=-2\left[c_{yp}(s)+c_{yp}^{*}(z)\right],B_{\eta}=2i\nu k_{u}(s-z)\nonumber \\
 & B_{y}=-B_{\alpha}c_{s}(s,z),B_{py}=B_{\alpha}\beta_{y}c_{c}(s,z),B_{\alpha}\equiv i\nu\frac{2K_{0}^{2}}{2+K_{0}^{2}}\alpha k_{u}\beta_{y}\nonumber \\
 & \text{where }c_{s}(s,z)\equiv\left(\sin(\frac{s}{\beta_{y}})-\sin(\frac{z}{\beta_{y}})\right),c_{c}(s,z)\equiv\left(\cos(\frac{s}{\beta_{y}})-\cos(\frac{z}{\beta_{y}})\right)
\end{align}

$\Phi_{x}$ and $\Phi_{y\eta}$ are quadratic polynomials in $x,p_{x}$
and in $\eta,y,p_{y}$respectively. Their coefficients are functions
of $s,z$ only. By linear transformation, they can be transformed
into diagonal quadratic form. Hence $I_{x}(s,z),I_{y\eta}(s,z)$ are
gaussian integrals. In Appendix II we give a brief description of
the process of transforming to gaussian integration. The result is

\begin{align}
 & I_{x}=\frac{2\pi\sigma_{px}\sigma_{x}}{\sqrt{(D_{x}D_{px}-\sigma_{x}^{2}\sigma_{px}^{2}B_{xp}^{2})}}\label{eq:Ixyeta}\\
 & I_{y\eta}=\frac{\left(2\pi\right)^{\frac{3}{2}}}{\sqrt{D_{py}}}\frac{\sigma_{\eta}\sigma_{y}\sigma_{py}}{\sqrt{D_{\eta y}}}\exp\left(\frac{N_{\eta y}}{\ensuremath{D_{\eta y}}}\sigma_{\eta}^{2}\sigma_{y}^{2}\right)\nonumber 
\end{align}

where $D_{x},D_{px},D_{py},B_{xp}$ are given in Eq.(\ref{eq:phix},\ref{eq:phietay})
and

\begin{align}
 & D_{\eta y}=1+\left(\sigma_{y}^{2}+D^{2}\sigma_{\eta}^{2}\right)\left(2A_{y}-\frac{1}{\sigma_{y}^{2}}-\frac{\sigma_{py}^{2}B_{yp}^{2}}{D_{py}}\right)\nonumber \\
 & N_{\eta y}=C_{ss}c_{s}^{2}(s,z)+C_{0}+C_{s}c_{s}(s,z)+C_{sc}c_{s}(s,z)c_{c}(s,z)+C_{cc}c_{c}^{2}(s,z)+C_{c}c_{c}(s,z)\label{eq:NDetay}
\end{align}

The sinusoidal functions $c_{s}(s,z)$,$c_{c}(s,z)$ are given by
Eq.(\ref{eq:phietay}). Their coefficients in $N_{\eta y}(s,z)$ are
also functions of $s,z$:

\begin{align}
\text{} & C_{ss}=B_{\alpha}^{2}A_{\eta},C_{0}=B_{\eta}^{2}\left(A_{y}-\frac{B_{yp}^{2}}{4A_{py}}\right),C_{s}=B_{\eta}B_{\alpha}B_{\eta y}\nonumber \\
 & C_{sc}=\frac{B_{yp}B_{\alpha}^{2}\beta_{y}A_{\eta}}{A_{py}},C_{cc}=\frac{B_{\alpha}^{2}\beta_{y}^{2}}{A_{py}}\left(A_{\eta}A_{y}-\frac{1}{4}B_{\eta y}^{2}\right),C_{c}=\frac{B_{\eta}B_{\alpha}B_{\eta y}B_{yp}\beta_{y}}{2A_{py}}\label{eq:Coff}
\end{align}

With these provisions, we find the gain $G$ as a double integral
over $s,z$. First substitute Eq.(\ref{eq:NDetay}) into Eq.(\ref{eq:I1e})
to find $I_{1e}$, then substitute $I_{1e}$, $\left(\int\mid A^{(0)}(\boldsymbol{\phi})\mid^{2}d\boldsymbol{\phi}\right)^{-1}$from
Eq.(\ref{eq:Ireq}), and the constant$c_{h}$ in Eq.(\ref{eq.ch})
into Eq.(\ref{G2}), with $\nu\approx h$, we finally have

\begin{align}
\text{} & G=-\left(\frac{hK^{2}[JJ]_{h}^{2}}{\lambda_{u}}\right)\left(\frac{I}{I_{A}}\frac{\pi^{2}}{\gamma^{3}}\right)\left(2I_{3D}\right)\frac{\text{\ensuremath{2\pi w_{x}w_{y}}}}{\lambda^{2}},\nonumber \\
 & \text{with }I_{3D}=\intop_{-L/2}^{L/2}\intop_{-L/2}^{L/2}dsdz\frac{-i(s-z)\exp(i\Delta\nu k_{u}(s-z))}{\sqrt{(D_{x}D_{px}-\sigma_{x}^{2}\sigma_{px}^{2}B_{xp}^{2})}\sqrt{D_{py}}\sqrt{D_{\eta y}}}\exp\left(\frac{N_{\eta y}}{\ensuremath{D_{\eta y}}}\sigma_{\eta}^{2}\sigma_{y}^{2}\right)\frac{1}{D_{R}(s,z)}\label{eq:Gfinal}
\end{align}

where $D_{\eta y},N_{\eta y}$ are given by Eqs.(\ref{eq:NDetay},\ref{eq:Coff}).
The coefficients in $D_{\eta y},N_{\eta y}$ are expressed by the
coefficients $A_{\eta},A_{y},A_{py}$ ,$B_{\eta},B_{\alpha},B_{\eta y},B_{yp},B_{xp}$and
$D_{x},D_{px},D_{py}$ of the polynomial $\Phi_{x},\Phi_{y\eta}$
in the gaussion integral, as given by Eq.(\ref{eq:phix},\ref{eq:phietay}).
The expressions in these coefficients, $c_{x}(s),c_{xp}(s),c_{px}(s),c_{y}(s),c_{yp}(s),c_{py}(s)$,
are given in Eq.(\ref{eq:Ireq}). The sinusoldal functions $c_{s}(s,z),c_{c}(s,z)$
in $N_{\eta y}$ are given in Eq.(\ref{eq:phietay}).

Because of effect of the betatron motion, the structure of the factors
in $I_{3D}$ is more complicated than the corresponding double integral
in \citep{lindberg,ysLi}. However, the numerical calculation of the
double integral is simple, so it is appropriate for optimization.

As a check, in 1D limit, $D_{x}=D_{px}=D_{py}=D_{\eta y}=1$,$B_{xp}=0$,$D_{R}(s,z)=z_{Rx}z_{Ry}=\frac{\pi^{2}w_{x}^{2}w_{y}^{2}}{\lambda^{2}}$.
The radiation beam size is the same as the electron beam size $\sigma_{rx}=w_{x}/2=\sigma_{x}$,$\sigma_{ry}=w_{y}/2=\sigma_{y}$
. If energy spread is negligible, $\sigma_{\eta}=0$, we have

\begin{align*}
 & I_{3D}=\left(\intop_{-L/2}^{L/2}\intop_{-L/2}^{L/2}dsdz\left(-i(s-z)\exp(i\Delta\nu k_{u}(s-z))\right)\right)\frac{\lambda^{2}}{\pi^{2}w_{x}^{2}w_{y}^{2}}\\
 & =-L^{3}\frac{1}{2}\frac{d}{dx}\left(\frac{\sin\Phi}{\Phi}\right)_{\Phi=\Delta\nu k_{u}L/2}^{2}\left(\frac{\lambda^{2}}{\pi^{2}w_{x}^{2}w_{y}^{2}}\right)
\end{align*}

Then $G$ becomes the $G_{1D}$ given by Eq.(\ref{eq:G1D}).

\section*{IV An Example of Gain Calculation}

The Eq.(\ref{eq:Gfinal}) is the main result of this paper. Our goal
is to apply this formula to explore the possibility of a hard x-ray
FEL for a light source at energy as low as 3GeV, and in particular,
to find the required electron beam quality and undulator for an upgrade
of NSLSII to drive a hard x-ray FEL to help to study if it is possible.

\begin{figure}[b]
\includegraphics[width=0.24\textwidth]{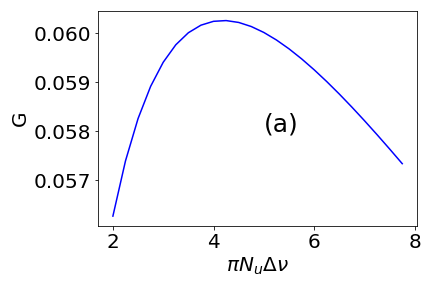}\includegraphics[width=0.25\textwidth]{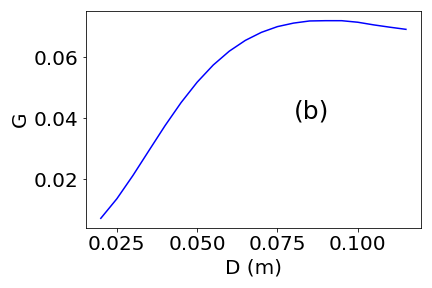}\includegraphics[width=0.25\textwidth]{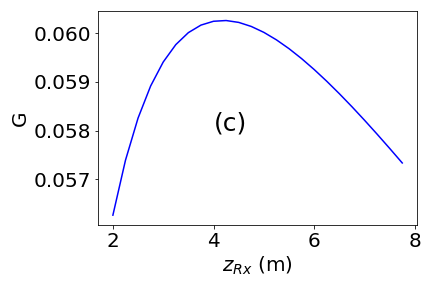}\includegraphics[width=0.25\textwidth]{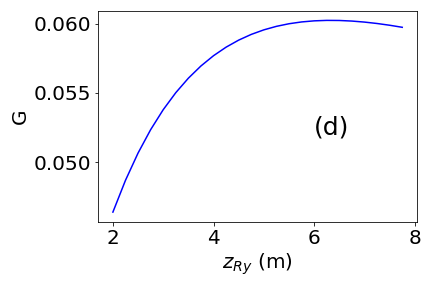}\vspace{0.0em}

\includegraphics[width=0.25\textwidth]{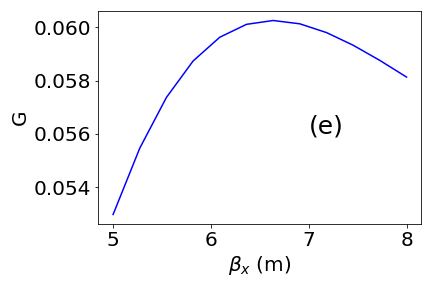}\includegraphics[width=0.25\textwidth]{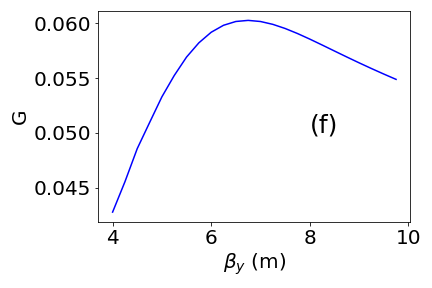}\includegraphics[width=0.25\textwidth]{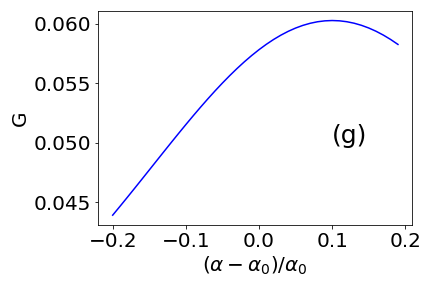}\includegraphics[width=0.25\textwidth]{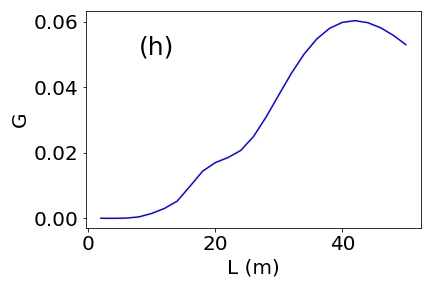}

\caption{G vs. (a)$\Delta\nu$, \ \ (b)D, \ \ (c)$z_{Rx}$ ,\ \ (d)$z_{Rx}$,
\ \ (e)$\beta_{x}$, \ \ (f)$\beta_{y}$, \ \  (g)($\alpha-\alpha_{0})/\alpha_{0}$,
\ \ (h) L}
\vspace{-1.0em}
\end{figure}

\begin{table}[t]
\begin{tabular}[t]{|c|c|c|c|c|c|c|c|c|c|c|c|c|c|c|c|}
\hline 
E (GeV) & I (A) & $\lambda_{u}$(cm) & h & G(\%) & $K_{0}$ & $\epsilon_{x}$ (pm) & $\epsilon_{y}$ (pm) & D (m) & $\beta_{x}$(m) & $\beta_{y}$(m) & $z_{Rx}$(m) & $z_{Ry}$(m) & $\alpha$ ($m^{-1})$ & $\sigma_{x}$($\mu m)$ & $\sigma_{y}$($\mu m)$\tabularnewline
\hline 
\hline 
3 & 73 & 2.26 & 5 & 6.0 & 1.29 & 80 & 1.7 & 0.05(V) & 6.6 & 6.9 & 4.0 & 6.3 & 49 & 23 & 3.5\tabularnewline
\hline 
\end{tabular}

\caption{Parameters for maximum $G$}
\end{table}
First, taking collective effects into account, we assume a 3GeV FEL
at 0.12nm, approximate the bunch as a flat top pulse, the current
$I=73A$ for a bunch length 180ps. For this example, we assume a local
coupling correction in the undulator section to minimize the local
vertical emittance and to blow up the vertical emittance in the rest
of the ring for mitigation of the intra-beam scattering. The revolution
period is about 2.6$\mu s$, so the bunch current is 5mA. As discussed
in the Section I, we plot the first factor $\frac{K^{2}[JJ]_{h}^{2}h}{\lambda_{u}}$
in the 1D gain formula as function of $\lambda_{u}$ in Fig.1a, and
plot the gap $g$ vs. $\lambda_{u}$ in Fig.1b. For $h=5$, as a compromise
between larger gain and gap, we choose $\lambda_{u}=2.26cm$ and $\frac{K^{2}[JJ]_{h}^{2}h}{\lambda_{u}}=4.96m^{-1}$,
with gap $g=7.5mm$, $K_{0}=1.29$.

We assume the emittance $\epsilon_{x}=80pm,\epsilon_{y}=1.7\,pm$,
energy spread $\sigma_{\eta}=10^{-3}$, and undulator length $L=42m$,
and scan the 6 variables: detuning $\Delta\nu$, horizontal dispersion
$D$, focusing beta functions $\beta_{x},\beta_{y},$the input radiation
Rayleigh ranges $z_{Rx},z_{Ry}$, and the transverse gradient $\alpha$
to find maximum gain. During the scan, we find the transverse gradient
$\alpha$ should be allowed to deviate from $\alpha_{0}$ (where $\alpha_{0}D=\frac{2+K_{0}^{2}}{K_{0}^{2}}$).
Actually, we find the optimized $\alpha$ close but larger than $\alpha_{0}$.

The plot of scan in the last scan cycle is given in Fig.3. The maximum
gain is 7\% in the last cycle for dispersion at 8 cm. However we limit
the vertical dispersion to 5 cm and the gain is 6.0 \%. In the subplot
of G vs. L the maximum gain is reached at $L=42m$. The paramters
for this setting are given in Table 1. In Table 1 the energy spread
is $\sigma_{\eta}=10^{-3}$, the undulator length is taken as $L=42m$,
the x-ray wavelength is $\lambda_{s}=0.12nm$, gap $g=7.5mm$. For
dispersion $D$, (V) is vertical dispersion.

\section*{V. GAIN OPTIMIZATION WITH COLLECTIVE EFFECTS}

As one can see in the previous section, XFELO requires quite a high
peak beam current and small emittance. The emittance of synchrotron
light sources is continuously reduced in past decades. Implementation
of the Multi-Bend Achromat (MBA) technology resulted in the development
of a new generation of synchrotrons with much lower emittance. Recently,
three new MBA-based rings have been commissioned \citep{tavares,torino,lin}
and few upgrade projects are being developed worldwide \citep{APS,steier,jiao,Karantzoulis,Diamond,shin}.
We consider an XFELO option for a lattice based on the recently developed
complex bend approach for the future low-emittance upgrade of NSLS-II
\citep{Plassard}. This lattice provides a horizontal emittance of
25 pm at a beam energy of 3 GeV fitting the present NSLS-II tunnel
with a circumference of 792 m. We propose to place the 42-m long XFELO
undulator in a straight section with a vertical dispersion bypassing
three out of 30 achromat cells.

However, collective effects of beam dynamics significantly affect
electron beam parameters in medium-energy (3-4 GeV) storage rings
because the beams are small in all three dimensions and the particle
density in the bunch is quite high. The main adverse effect impeding
the achievement of the required combination of beam parameters is
intra-beam scattering (IBS). To mitigate the collective effects, higher-harmonic
RF cavities are used for bunch lengthening. The other strong intensity-dependent
effect is the bunch lengthening due to potential well distortion by
the longitudinal impedance of the vacuum chamber.

\begin{figure}[b]
\includegraphics[width=0.33\textwidth]{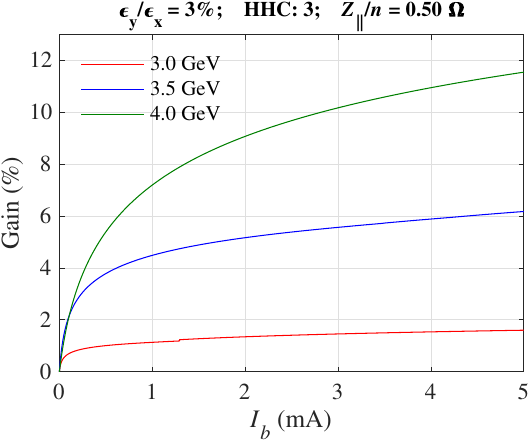}\vspace{-1.0em}

\caption{FEL gain optimized taking collective effects into account}
\vspace{-1.0em}
\end{figure}

For a realistic assessment of a ring-based XFELO, we carried out multi-parameter
optimization of the FEL gain Eq.\ref{eq:Gfinal} assuming the vertical
dispersion of 5 cm, emittance, energy spread, and bunch length determined
by the lattice model taking into account the effect of IBS together
with the impedance-driven bunch lengthening and higher-harmonic cavities.
For the low-emittance synchrotrons, the light-generating insertion
devices make a major contribution to the total energy loss per turn
$U_{0}$ determining the radiation damping, so we include them into
the lattice model. We optimized the detuning $\Delta\nu$, beta functions
$\beta_{x},\beta_{y}$, the input radiation Rayleigh ranges $z_{Rx},z_{Ry}$,
and the transverse gradient $\alpha$ to find the maximum gain.

We applied the high-energy approximation of the IBS theory\citep{bane}.
The equilibrium emittance $\mbox{\ensuremath{\varepsilon_{x,y}}}$
and relative energy spread $\mbox{\ensuremath{\sigma_{p}}}$ are expressed
as

\begin{align*}
 & \varepsilon_{x,y}=\frac{\varepsilon_{x0,y0}}{1-\tau_{x,y}/T_{x,y}}\ ,\qquad\sigma_{p}^{2}=\frac{\sigma_{p0}^{2}}{1-\tau_{p}/T_{p}}\ ,
\end{align*}

where $\varepsilon_{x0,y0}$ and $\sigma_{p0}$ are the emittance
and energy spread at zero beam current; $\ensuremath{\tau_{x}}$,
$\tau_{y}$, and $\tau_{p}$ are the radiation damping times; $T_{xy}$,
and $T_{p}$ are the IBS growth times:

\[
\frac{1}{T_{p}}\simeq\frac{r_{0}^{2}cN}{32\gamma^{3}\varepsilon_{x}\varepsilon_{y}\sigma_{s}\sigma_{p}^{2}}\left(\frac{\varepsilon_{x}\varepsilon_{y}}{\left<\beta_{x}\right>\left<\beta_{y}\right>}\right)^{\!\!1/4}\ln\frac{\left<\sigma_{y}\right>\gamma^{2}\varepsilon_{x}}{r_{0}\left<\beta_{x}\right>}\ ,
\]

\[
\frac{1}{T_{x,y}}\simeq\frac{\sigma_{p}^{2}\left<\mathscr{H}_{x,y}\right>}{\varepsilon_{x,y}}\frac{1}{T_{p}}\ ,
\]

$r_{0}$ is classical electron radius, $\sigma_{s}$ is the r.m.s.
bunch length, $\sigma_{y}$ is the vertical beam size, $\mathscr{H}_{x,y}$
is the lattice function Eq.\ref{lattice_function}

\begin{equation}
\mathscr{H}=\beta_{x}\eta_{x}'^{2}+2\alpha_{x}\eta_{x}\eta_{x}'+\gamma_{x}\eta_{x}^{2}\ ,\label{lattice_function}
\end{equation}

$\beta_{x}$ is the amplitude function of betatron oscillation (beta
function), $\alpha_{x}\equiv-\beta_{x}'/2$, $\gamma_{x}\equiv\frac{1+\alpha_{x}^{2}}{\beta_{x}}$;
$\eta_{x}$ and $\eta_{x}'$ is the dispersion function and its derivative,
respectively. As one can see, the IBS strongly depends on the beam
energy, so its effect is not so significant for high-energy rings.

\textcolor{black}{The bunch lengthening caused by the beam interaction
with the longitudinal impedance was calculated using the modified
Zotter equation\citep{zotter} in differential form. The effect of
higher-harmonic cavities was simply modeled by a multiplication of
the zero-intensity bunch length by a factor of 3.}

\textcolor{black}{As a result of the optimization, Fig.4 shows the
FEL gain as a function of single-bunch current for the beam energy
of 3GeV, 3.5GeV, and 4GeV, and for an extremely low coupling of 3\%.}

\textcolor{black}{With the constant coupling, the 3-GeV XFELO does
not look realistic because the gain is too small but the energy increase
results in a feasible gain of 6\%, especially for 3.5GeV to 4GeV.}
\begin{figure}[t]
\textcolor{black}{\includegraphics[width=0.33\textwidth]{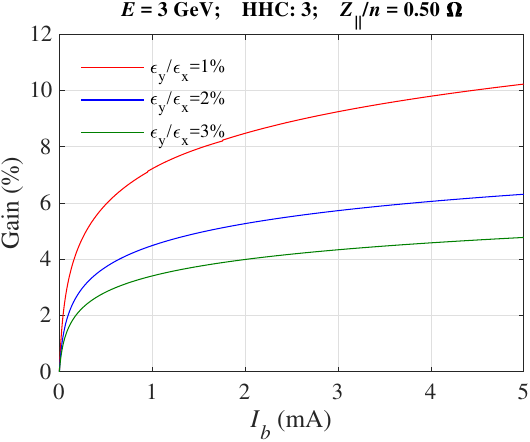}\vspace{-1.0em}}

\textcolor{black}{\caption{FEL gain optimized with a local coupling correction}
\vspace{-1.0em}}
\end{figure}

\textcolor{black}{However, if we implement a local coupling correction
to reduce the vertical beam size only in the undulator section with
coupling of 2\% while keeping 100\% coupling everywhere else in the
ring, the XFELO looks feasible with gain reaching 6\% even for 3$\sim$GeV
energy, see Fig.5.}

\section*{VI. Summary}

We developed a gain formula for a hard x-ray FEL in the medium energy
range between 3 and 4.5GeV so that we do not need to take ``no focusing
approximation'' in the calculation, in the hope this can be of use
in exploring the possibility of x-ray FEL in this energy range. The
formula allows gain calculation with harmonic lasing.

We present an example. The example indicates hard x-ray FEL in 3GeV
is feasible, even though it sets rather challenging conditions for
the storage ring.

\section*{Appendix I Outline of derivation of FEL gain in small signal, low
gain regime}

The FEL gain is derived from the coupled Maxwell-Vlasov equations,
which initially was developed for high gain FEL theory \citep{wang,prA,PRL,HGHG}.
Later it is used for small gain FEL in \citep{lindberg,ysLi,kim4}.
We outline the gain formula derivation here as given in \citep{kim4}
in order to clarify the inclusion of the TQU, the dispersion, and
we pay attenion to the harmonic number $\nu\approx h$ in this paper.
Many of the notations here have been introduced in Section II when
we introduce the gain formula. The Maxwell equation in the notation
here is

\begin{equation}
\begin{aligned} & \begin{split}(\frac{\partial}{\partial z}-i\Delta\nu k_{u}\end{split}
+\frac{i}{2k}\nabla_{\bot}^{2})a_{\nu}(x,z)=-g_{n}\int d\eta\int d\mathbf{p}\delta F_{\nu}(z,\theta,\eta,\mathbf{x},\mathbf{p})\end{aligned}
\label{eq:maxwell}
\end{equation}

where $g_{n}=\frac{eK[JJ]_{h}n_{0}}{4\epsilon_{0}\gamma\omega_{1}}$,
and $n_{0}F(z,\theta,\eta,\mathbf{x},\mathbf{p})=k_{1}\sum\delta(\eta-\eta_{j})\delta(\mathbf{x}-\mathbf{x}_{j})\delta(\mathbf{p}-\mathbf{p}_{j})\delta(\theta-\theta_{j})$
is the electron beam distribution function, $\theta=(k_{1}+k_{u})z-\omega_{1}\overline{t}$
is the micro-bunching phase, $\overline{t}$ is the time of electron
passing through $z$ averaged over a period of the undulator. The
function $F$ is separated into two parts $F=\overline{F}+\delta F$,
$\overline{F}$ is the averaged smooth electron beam background distribution
in Eq.(\ref{eq:F}), $\delta F$ is the deviation from $\overline{F}$
including shot noise in the beam and the micro-bunching due to FEL
interaction. $\delta F_{\nu}(z,\theta,\eta,\mathbf{x},\mathbf{p})$
is the Fourier transfrom of $\delta F$:

\[
\delta F_{\nu}(\eta,\mathbf{x},\mathbf{p};z)=\frac{1}{\sqrt{2\pi}}\int e^{i\nu\theta}\delta F(z,\theta,\eta,\mathbf{x},\mathbf{p})d\theta
\]

The Vlasov equation is essentially the Liouville theorem applied to
perturbation theory in a small signal regime. After Fourier transfrom
from $\theta$ to $\nu$, it is in the form

\begin{equation}
\begin{split}(\frac{\partial}{\partial z}\end{split}
+\dot{\theta}\frac{\partial}{\partial\theta}+\dot{\mathbf{x}}\frac{\partial}{\partial\mathbf{x}}+\dot{\mathbf{p}}\frac{\partial}{\partial\mathbf{p}})\delta F_{\nu}(\eta,\mathbf{x},\mathbf{p};z)=-h_{h}a_{\nu}(\mathbf{x},z)\frac{\partial}{\partial\eta}\bar{F}(\eta,\mathbf{x},\mathbf{p};z)\label{eq:vlasov}
\end{equation}

where the Fourier transformed energy equation is $\frac{1}{\sqrt{2\pi}}\int\dot{\eta}e^{i\nu\theta}d\theta=-h_{n}a_{\nu}(x,y,z)$,
with $h_{h}=\frac{e\omega_{1}K[JJ]_{h}}{2mc^{2}\gamma^{2}}$, $\dot{x},\dot{p}$
are given by betatron motion Eq.(\ref{eq:betaeq}), and the phase
equation is

\begin{align*}
 & \dot{\theta}=\frac{d\theta}{dz}=k_{u}\left(2\eta-\frac{K_{0}^{2}}{1+\frac{K_{0}^{2}}{2}}\alpha y\right)-\frac{k_{1}}{2}w(\mathbf{x},\mathbf{p})\\
 & \text{where }w(\mathbf{x},\mathbf{p})\equiv p_{x}^{2}+k_{\beta x}^{2}x^{2}+p_{y}^{2}+k_{\beta y}^{2}y^{2}
\end{align*}

To solve the coupled Maxwell-Vlasov equations with two unknowns $a_{\nu}(x,z),\delta F_{\nu}$,
we eliminate the unknown $\delta F_{\nu}$ by first solving the Vlasov
equation Eq.(\ref{eq:vlasov}) to express $\delta F_{\nu}$ in terms
of $a_{\nu}(x,z)$. Treating this linear partial differential equation
as a one variable linear ordinary differential equation with $\delta F_{\nu}$
as a function of $z$, using the ``method of variation of constants'',
the result is

\begin{align}
 & \delta F_{\nu}(\eta,\mathbf{x},\mathbf{p};z)=\delta F_{\nu}^{(0)}(\eta,\mathbf{x},\mathbf{p};z)-h_{n}\intop_{0}^{z}dsa_{\nu}(x_{0},s))\left(\frac{\partial}{\partial\eta}\overline{F}(\eta,\mathbf{x_{0}(\mathbf{x},\mathbf{p}},z;s),\mathbf{p_{0}(\mathbf{x},\mathbf{p}},z;s);s)\right)\nonumber \\
 & \times\exp\left(-2i\nu\eta k_{u}(s-z)+i\frac{\nu k_{1}}{2}\intop_{z}^{s}dz_{1}w(\mathbf{x_{0}(\mathbf{x},\mathbf{p}},z;z_{1}),\mathbf{p_{0}(\mathbf{x},\mathbf{p}},z;z_{1}))+i\nu k_{u}\frac{K_{0}^{2}}{1+\frac{K_{0}^{2}}{2}}\alpha\intop_{z}^{s}dz_{1}y_{0}\right)\label{eq:solF}
\end{align}

where $\delta F_{\nu}^{(0)}(\eta,\mathbf{x},\mathbf{p};z)$ is the
solution without FEL interaction, related to the spontaneous radiation
and will be neglected in the small gain calculation, and $\mathbf{x_{0}(\mathbf{x},\mathbf{p}},z;s),\mathbf{p_{0}(\mathbf{x},\mathbf{p}},z;s)$
are solution of the betatron motion Eq.(\ref{eq:betaeq}) with initial
conditon such that at $s=z$, $\mathbf{x_{0}(\mathbf{x},\mathbf{p}},z;s=z)=\mathbf{x}$
and $\mathbf{p_{0}(\mathbf{x},\mathbf{p}},z;s=z)=\mathbf{p}$.

Inserting Eq.(\ref{eq:solF}) into the Maxwell equation Eq.(\ref{eq:maxwell}),
and neglecting the first term which contributes to the spontaneous
radiation, we get the field equation for $a_{\nu}(x,z)$

\begin{align}
 & (\frac{\partial}{\partial z}-i\Delta\nu k_{u}+\frac{i}{2k}\nabla_{\bot}^{2})a_{\nu}(\mathbf{x},z)=-g_{n}h_{n}\int d\eta\int d\mathbf{p}\intop_{0}^{z}ds\overline{F}(\eta,\mathbf{x_{0}(}s),\mathbf{p_{0}(}s);s)\frac{\partial}{\partial\eta}a_{\nu}(\mathbf{x_{0}(}s),s)\nonumber \\
 & \times\exp\left(-2i\nu\eta k_{u}(s-z)+i\frac{k}{2}\intop_{z}^{s}dz_{1}w(\mathbf{x_{0}(}z_{1}),\mathbf{p_{0}(}z_{1}))+i\nu k_{u}\frac{K_{0}^{2}}{1+\frac{K_{0}^{2}}{2}}\alpha\intop_{z}^{s}dz_{1}y_{0}(z_{1})\right)\label{eq:fiedeq}
\end{align}

We abbreviate $\mathbf{x_{0}(\mathbf{x},\mathbf{p}},z;s)$ as $\mathbf{x_{0}(}s)$,
and similarly for $\mathbf{p_{0}(}s)$. $\mathbf{\mathbf{x},\mathbf{p}},z$
in the argument is implied, and $k=\nu k_{1}$. The version of this
equation with the transverse gradient $\alpha=0$ has been used as
dispersion relation \citep{PRL} to derive the gain length formula
including the effect of finite emittance, diffraction, and betatron
focusing for the development in the high gain regime for the first
time.

Later this equation was applied to solve for small gain regime \citep{lindberg,ysLi,kim4}.
Applying the transform $\frac{1}{\lambda^{2}}\intop d\mathbf{x}e^{ik\left(x\phi_{x}+y\phi_{y}\right)}$in
Eq.(\ref{eq:Aanular}) to both sides of the Eq.(\ref{eq:fiedeq})
converts it to a field equation for $A_{\nu}(\phi_{x},\phi_{y};z)$

\begin{align}
 & (\frac{\partial}{\partial z}-i\Delta\nu k_{u}-\frac{ik}{2}\boldsymbol{\phi}^{2})A_{\nu}(\boldsymbol{\phi};z)=b(z)\nonumber \\
 & \text{where }b(z;\phi)=-\frac{1}{\lambda^{2}}g_{n}h_{n}\int d\eta d\mathbf{p}\intop d\mathbf{x}\intop d\boldsymbol{\phi}'\exp(\Delta\nu k_{u}z+i\frac{k}{2}\boldsymbol{\phi}^{2}z)\overline{F}(\eta,\mathbf{x},p;z)\label{eq:fiedangular}\\
 & \times\frac{\partial}{\partial\eta}e^{ik\mathbf{x}\boldsymbol{\phi}}\exp(-i\intop_{0}^{z}dz_{1}\xi_{\nu}(\boldsymbol{\phi},\eta,\mathbf{x},\mathbf{p};z_{1}))\intop_{-L/2}^{z}ds_{1}e^{-ik\mathbf{x}\boldsymbol{\phi}'}\exp(i\intop_{0}^{s_{1}}dz_{1}\xi_{\nu}(\boldsymbol{\phi}',\eta,\mathbf{x},\mathbf{p};z_{1}))A_{\nu}(\boldsymbol{\phi}',0)\nonumber \\
 & \text{and }\xi_{\nu}(\phi',\eta,\mathbf{x},\mathbf{p};z_{1})=(\Delta\nu-2\nu\eta)k_{u}+\nu k_{u}\frac{K_{0}^{2}}{1+\frac{K_{0}^{2}}{2}}\alpha y_{0}(z_{1})+\frac{k}{2}w(\mathbf{x_{0}(}z_{1}),\mathbf{p_{0}(}z_{1})-\boldsymbol{\phi})
\end{align}

The input radiation $A_{\nu}^{(0)}(\phi_{x},\phi_{y},z)$ in Eq.(\ref{eq:Aanular}),
the gaussian beam, is the solution of this equation with $b(z)$ set
to zero. When substituting $A_{\nu}^{(0)}(\phi_{x},\phi_{y},0)$ as
$A_{\nu}(\boldsymbol{\phi}',0)$ into Eq.(\ref{eq:fiedangular}) first-order
perturbation, the equation is considered as a linear ordinary differential
equation with $z$ as variable and $b(z)$ as the inhomogeneous term,
the solution at the end $z=L/2$ is

\begin{align}
 & A_{\nu}(\boldsymbol{\phi};L/2)=A_{\nu}^{(0)}(\phi;L/2)+A_{\nu}^{(1)}(\phi,L/2)\nonumber \\
 & \text{with }A_{\nu}^{(0)}(\phi;L/2)=\exp(i\Delta\nu k_{u}L/2+\frac{ik}{2}\phi^{2}L/2)A_{\nu}^{(0)}(\phi',0)\nonumber \\
 & A_{\nu}^{(1)}(\boldsymbol{\phi})=-\frac{1}{\lambda^{2}}g_{h}h_{h}\exp(i\Delta\nu k_{u}L/2+\frac{ik}{2}\phi^{2}L/2)\int d\eta dp\intop dx\intop d\boldsymbol{\phi}'\overline{F}(\eta,\mathbf{x},\mathbf{p};0)\label{eq:Anu1}\\
 & \times\frac{\partial}{\partial\eta}\intop_{-L/2}^{L/2}dse^{ik\mathbf{x}\boldsymbol{\phi}}\exp(-i\intop_{0}^{s}dz_{1}\xi_{\nu}(\boldsymbol{\phi},\eta,\mathbf{x},\mathbf{p};z_{1}))\intop_{-L/2}^{s}ds'e^{-ik\mathbf{x}\boldsymbol{\phi}'}\exp(i\intop_{0}^{s'}dz_{1}\xi_{\nu}(\boldsymbol{\phi}',\eta,\mathbf{x},\mathbf{p};z_{1}))A_{\nu}(\phi',0)\nonumber 
\end{align}

here $\overline{F}(\eta,\mathbf{x},p;z)$ in Eq.(\ref{eq:fiedangular})
has been replaced by $\overline{F}(\eta,\mathbf{x},p;0)$ because
it is independent of $z$. In Eq.(\ref{eq:Anu1}), and in Eq.(\ref{eq:general gain})
of Section II the $\mathbf{x}_{0}(s;\mathbf{x},\mathbf{p})$,$\mathbf{p}_{0}(s;\mathbf{x},\mathbf{p})$
in $\xi_{\nu}(\boldsymbol{\phi}',\eta,\mathbf{x},\mathbf{p};z_{1})$
are solutions of the equations of betatron motion Eq.(\ref{eq:betaeq})
with the initial condition.

The gain is defined as

\begin{align}
 & G=\frac{\int\left(\mid A_{\nu}(\boldsymbol{\phi};L/2)\mid^{2}-\mid(A_{\nu}^{(0)}(\boldsymbol{\phi};L/2)\mid^{2}\right)d\phi}{\int\mid(A_{\nu}^{(0)}(\boldsymbol{\phi};L/2)\mid^{2}d\boldsymbol{\phi}}\label{eq:G}\\
 & \approx\frac{\int\left(A_{\nu}^{(0)*}(\boldsymbol{\phi};L/2)A_{\nu}^{(1)}(\boldsymbol{\phi},L/2)+c.c.\right)d\phi}{\int\mid(A_{\nu}^{(0)}(\boldsymbol{\phi};L/2)\mid^{2}d\boldsymbol{\phi}}\nonumber 
\end{align}

where the term of second power of $\frac{1}{\lambda^{2}}g_{n}h_{n}\text{ }$
is neglected. Substitute Eq.(\ref{eq:Anu1}) in Eq.(\ref{eq:G}),
the result is the gain formula Eq.(\ref{eq:general gain}).

\section*{Appendix II Gaussian Integral of several variables}

We brief the calculation of the 3 variable gaussian integrals $I_{y\eta}$
in Eq.(\ref{eq:Ixyeta}) of Section III. The integral $I_{x}$ is
a 2-variable gaussian, and can be considered as a simplified version
of $I_{y\eta}$. The exponent in $I_{y\eta}(s,z)=\int\int d\eta dydp_{y}\exp(-\Phi_{y\eta})$
is

\[
\Phi_{y\eta}=A_{\eta}\eta^{2}+A_{y}y^{2}+A_{py}p_{y}^{2}+B_{\eta y}y\eta+B_{yp}y\eta+B_{\eta}\eta+B_{y}y+B_{py}p_{y}
\]

where the coefficients are given by Eq.(\ref{eq:phietay}). The first
step is to transform the coefficients of the 3 quadratic terms to
1 by a transform $A_{\eta}\implies A_{\eta}/\sqrt{A_{\eta}}$ $y\implies y/\sqrt{A_{y}},p_{y}\implies p_{y}/\sqrt{A_{py}}$
so 
\begin{align*}
 & I_{y\eta}=\frac{1}{\sqrt{A_{\eta}A_{py}A_{y}}}\int\int d\eta dydp_{y}\exp(-\Phi_{y\eta1})\\
 & \Phi_{y\eta1}=\eta^{2}+p_{y}^{2}+y^{2}+dy\eta+cyp_{y}+a\eta+by+gp_{y}
\end{align*}

where $a=\frac{B_{\eta}}{\sqrt{A_{\eta}}}$,$b=\frac{-B_{\alpha}}{\sqrt{A_{y}}}\left(\sin(\frac{s}{\beta_{y}})-\sin(\frac{z}{\beta_{y}})\right),B_{\alpha}\equiv i4\nu\zeta_{0}\alpha k_{u}\beta_{y}$,$c=\frac{B_{yp}}{\sqrt{A_{py}A_{y}}}$,$d=\frac{B_{\eta y}}{\sqrt{A_{\eta}A_{y}}}$,$g=\frac{B_{\alpha}\beta_{y}}{\sqrt{A_{py}}}\left(\cos(\frac{s}{\beta_{y}})-\cos(\frac{z}{\beta_{y}})\right)$.

Next, we shift the origin to the maximum of $\Phi_{y\eta1}$at $\eta_{0},p_{y0},y_{0}$
by a transform $\eta\implies\eta+\eta_{0},p_{y}\implies p_{y}+p_{y0},y\implies y+y_{0}$,
which is found by solving 3 linear equations. The result is

\begin{align*}
 & I_{y\eta}=\frac{1}{\sqrt{A_{\eta}}\sqrt{A_{py}}\sqrt{A_{y}}}\exp(-\frac{4b^{2}+a^{2}(4-c^{2})-4abd-4bcg+(4-d^{2})g^{2}+2acdg}{4(4-c^{2}-d^{2})})\int\int d\eta dydp_{y}\exp(-\Phi_{y\eta2})\\
 & \Phi_{y\eta2}=X^{T}.m.X=\eta^{2}+p_{y}^{2}+(\eta d+cp_{y})y+y^{2}\\
 & \text{where }X=\left(\begin{array}{c}
\eta\\
p_{y}\\
y
\end{array}\right),m=\left(\begin{array}{ccc}
1 & 0 & \frac{d}{2}\\
0 & 1 & \frac{c}{2}\\
\frac{d}{2} & \frac{c}{2} & 1
\end{array}\right)
\end{align*}

$\Phi_{y\eta2}$ is in quadratic form and can be transformed into
diagonal quadratic form using the eigenvectors $V$ of the matrix
$m$

\[
V=\left(\begin{array}{ccc}
-\frac{c}{d} & -\frac{d}{\sqrt{c^{2}+d^{2}}} & \frac{d}{\sqrt{c^{2}+d^{2}}}\\
1 & -\frac{c}{\sqrt{c^{2}+d^{2}}} & \frac{c}{\sqrt{c^{2}+d^{2}}}\\
0 & 1 & 1
\end{array}\right)
\]
Now apply transform:$X\implies VX$ with determinant $\det(V)=\frac{2\sqrt{c^{2}+d^{2}}}{d}$,
we find

\begin{align*}
 & I_{y\eta}=\frac{1}{\sqrt{A_{\eta}A_{py}A_{y}}}\frac{2\sqrt{c^{2}+d^{2}}}{d}\exp\left(\frac{4b^{2}+a^{2}(4-c^{2})-4abd-4bcg+(4-d^{2})g^{2}+2acdg}{4(4-c^{2}-d^{2})}\right)\\
 & \times\int\int d\eta dydp_{y}\exp(-(1+\frac{c^{2}}{d^{2}})\eta^{2}-\left(2-\sqrt{c^{2}+d^{2}}\right)p_{y}^{2}-\left(2+\sqrt{c^{2}+d^{2}}\right)y^{2})
\end{align*}

The integrals over $\eta,y,p_{y}$ are separately carried out, and
finanly with $a,b,c,d,g$ substituted and rearranged, the result is
Eq.(\ref{eq:Ixyeta}).

\end{document}